\documentstyle[preprint,aps]{revtex}

\begin{document}

\preprint{ORNL-CTP-95-03 and ASHEP-95-01}

\draft

\title{ Effect of $q\bar q$ Initial-state Interaction on Dilepton
  Emission Rate from Quark-Gluon Plasma}

\author{ M. G.-H. Mostafa$^{1,2}$, Cheuk-Yin Wong$^1$, Lali
  Chatterjee$^3$, and Zhong-Qi Wang$^4$}

\address{ $^1$Oak Ridge National Laboratory, Oak Ridge, TN 37831}

\address{ $^2$Physics Department, Faculty of Science, Ain Shams
  University, Cairo, Egypt}

\address {$^3$Jadavpur University, Calcutta 700032, India}

\address {$^4$China Institute of Atomic Energy, Beijing, China}

\date{\today}

\maketitle

\begin{abstract}
  We calculate the dilepton production rate from a thermalized
  quark-gluon plasma in heavy-ion collisions at RHIC
  energies. Higher-order QCD corrections are included by using an
  analytical correction factor $K^{(i)}$, which takes into account
  the $q\bar q$ initial-state interactions.  We show that the
  analytic correction factor gives very good agreement with
  experimental Drell-Yan data and leads to large enhancement of
  the thermal dilepton emission rates. We compare the thermal
  dilepton yields with the expected production from open-charm
  decays and Drell-Yan background and assess the prospects of
  observing thermal dileptons from the quark-gluon-plasma at
  invariant masses of a few GeV.

\end{abstract}


\narrowtext


High-energy heavy-ion collisions have become the focus of intense
experimental and theoretical research in recent years because of
the possibility of producing hadron matter in a deconfined
quark-gluon plasma (QGP) state during such
collisions\cite{Won1}. Various signals have been proposed as a
probe of this new phase of nuclear matter. In particular,
electromagnetic signals comprising dileptons and photons, are
recognized as direct probes for the QGP phase\cite{Mul1,Kap,Ruu},
since they weakly interact with the hadronic medium in which they
are produced. The magnitudes of the dilepton yields depend on the
cross sections for the basic process producing the dileptons.

The dilepton signals indicative of the QGP phase are those
produced by the annihilation of $q\bar q$ pairs through virtual
photon intermediate states. To be meaningful signals, they must be
clearly delineated from other sources of dilepton pairs like
Drell-Yan process, open charm decays, vector meson decays, and so
on. The problems associated with these backgrounds have been
discussed in the literature \cite{Cle,Kaj1,Kaj2,Ruu,Lei,Maz} and
their contributions are known to vary with the invariant masses of
the dileptons and the kinematic regions under
consideration\cite{Mul1,Vog,Lou,Aki,Mul2}. For dilepton pairs in
the invariant mass region of $2$ to $7$ Gev, the charm dileptons
and DY dileptons are expected to be the major background to
detection of thermal dileptons from the quark-gluon-plasma
phase. We shall discuss our thermal dilepton predictions in the
light of these backgrounds.

    Thermal dileptons as well as Drell Yan pairs originate from
the electromagnetic annihilation of quark-antiquark pairs through
intermediate virtual photons.  While the surroundings that host
the processes for dilepton production from the thermal plasma or
the DY production during nuclear collisions are vastly different,
the basic process corresponding to the channel $q\bar q\to
\gamma^* \to l^+ l^-$ is the same for the two cases. For
comparative analyses, therefore, the dileptons should be corrected
similarly for the effects that modify the tree-level diagrams,
whether the dilepton source is the thermal plasma or the DY
mechanism.  For the latter processes, it has been already
generally recognized in perturbative QCD that the lowest-order
Feynman diagrams give only an approximate description
\cite{Kub80,Ham91,Fie89,Bro93,Gro86}. In order to bring the
lowest-order QCD predictions into agreement with experiment, the
lowest-order DY results must be multiplied by a phenomenological
$K$-factor, which has a magnitude of the order of $3$. The
$K$-factor can be accounted for in terms of higher-order QCD
corrections, the most important effects being the vertex
correction arising from the initial-state interactions of the $q
\bar q$, prior to their annihilation into the virtual photon
producing the dileptons\cite{Kub80,Ham91}.  These higher-order QCD
corrections are operative also for $q \bar q$ annihilation in the
QGP, and the dilepton production cross section in the QGP must be
similarly corrected.

 In this work we investigate dilepton production from the QGP in
nucleus-nucleus collisions at RHIC energies, by incorporating
these QCD corrections through an analytic correction factor
presented recently\cite{Lal}.  We compare our calculated rates
with those expected from open charm production and the DY process
in the energy range of a few GeV.  Our calculated values of open
charm dileptons exceed those of the DY process, in conformity with
recent findings \cite{Aki}. However, we show that the thermal
dilepton yield, computed with the QCD corrections can be
comparable and in some scenarios can be well above the background,
so that prospects for its observation at RHIC energies becomes
promising.

 We use the analytical correction factor for the basic reaction
processes involving $q$ and $\bar q$ \cite{Lal}, for the whole
range of relative energies.  For low energies, the correction
factor is obtained by studying the distortion of the wave function
by virtue of the $q$-$\bar q$ color potential arising from virtual
gluon exchange.  At high energies, the correction factor is chosen
to match well-known PQCD results which give good agreement with
the experimental correction factor for the initial- or final-state
interactions as applicable. An interpolation to join the
correction factor from low energies to these well-known
perturbative QCD correction factors at high energies is made by
following the procedure suggested by Schwinger \cite{Sch73}.

 The analytical correction $K$-factor for the annihilation or the
production of a $q\bar q$ pair with an invariant mass $M^2=s$ is
explicitly given by\cite{Lal}
\begin{eqnarray}
  \label{eq:150}
  K^{(i,f)}(q) ={ 2 \pi f^{(i,f)}(v) \over 1- \exp \{ {-2 \pi
  f^{(i,f)}(v)} \} } ( 1+ \alpha_{\rm eff}^2 ) \,,
\end{eqnarray}
where the flavor label $q$ in $K^{(i)}(q)$ is included to indicate
that $K^{(i)}$ depends on the quark mass $m_q$, the superscripts
$(i)$ and $(f)$ denote $q\bar q$ initial-state annihilation and
$q\bar q$ final-state production respectively.
\begin{eqnarray}
  f^{(i)}(v)= \alpha_{\rm eff} \biggl [ { 1 \over v } + v \biggl
  ( - 1 + { 1 \over 2 \pi^2} + { 5 \over 6} \biggr ) \biggr ] \,,
\end{eqnarray}
\begin{eqnarray}
  \label{eq:130}
  f^{(f)}(v)= \alpha_{\rm eff} \biggl [ { 1 \over v } + v \biggl
  ( - 1 + {3 \over 4 \pi^2} \biggr ) \biggr ] \,.
\end{eqnarray}
In the above equations, $v$ is the relative asymptotic velocity for
the quark and the antiquark in their center-of-mass system
\begin{eqnarray}
   \label{eq:65}
   v= {(s^2 - 4s m_q^2)^{1/2} \over s - 2 m_q^2 },
\end{eqnarray}
and $\alpha_{\rm eff}$ is the effective strong interaction
coupling constant related to the strong interaction coupling
constant $\alpha_{\rm s}$ by the color factor $C_f$
\begin{eqnarray}
  \alpha_{\rm eff}={C_f\alpha_{\rm s} } \,.
\end{eqnarray}
For dilepton production, the $q$ and $\bar q$ must be in the color
singlet state, i.e. $C_f=-{4\over 3}$, and the running coupling
constant is taken as \cite{Fie89,Bro93}
\begin{eqnarray}
  \alpha_{\rm s} = { 12 \pi \over (33 - 2 n_f ) \ln
    (M^2/\Lambda^2) }\,.
\end{eqnarray}

To demonstrate the validity of the correction factor, we apply it
first to the Drell-Yan case and compare with known experimental
data.

The lowest-order Drell-Yan distribution for
 nucleon-nucleon collisions is given by

\begin{eqnarray}
\label {eqn:DYLO}
  {d^2\sigma^{NN}_{DY} \over dMdy} = { 8 \pi\alpha^2 \over 3 s N_c M}
  \sum_{q=u,d,s} \biggl ({e_{q}\over e} \biggr )^2  [ q_q^a (x e^y)
\bar q_q^b (x e^{-y})  +  \bar q_q^a (x e^y)q_q^b (x e^{-y})]
\end{eqnarray}
where $x=M/ \sqrt{s}$ and $y$ is the rapidity.  $q_q^{a,b}(x e^y)$
and $\bar q_q^{a,b}(x e^y)$ refer to the quark and antiquark
distributions in the nucleons $a$ and $b$ respectively.

Correcting for initial-state color interaction we obtain

\begin{eqnarray}
\label{eqn:DYK}
  {d^2\sigma_{DY}^{NN} \over dMdy} = { 8 \pi\alpha^2 \over 3 s N_c
  M} \sum_{q=u,d,s} K^{(i)} (q)\biggl ({e_{q}\over e} \biggr )^2 [
  q_q^a (x e^y) \bar q_q^b (x e^{-y}) + \bar q_q^a (x e^y)q_q^b (x
  e^{-y})]
\end{eqnarray}

The DY distributions for equal-nuclei nucleus-nucleus collisions
are obtained from the nucleon-nucleon case by using

\begin{eqnarray}
  {d^2N^{AA}_{DY} \over dMdy} = {3 \over 4 \pi (r'_0)^2}\;
  A^{4/3}\; {d^2 \sigma^{NN}_{DY} \over dMdy} \,,
\end{eqnarray}
where $A$ is the atomic number of the colliding nuclei and
$r_0^{'} = 1.2$ fm.

Figure~\ref{fig1} shows the FNAL-605 experimental data
\cite{Sti93} for the differential cross section $s d^2\sigma/d
\sqrt\tau dy$ as a function of $\sqrt\tau=M/\sqrt{s}$ for
different rapidity $y$ intervals, compared to the corrected
cross-sections using the Duke and Owens structure functions with
$\lambda=0.2$ GeV \cite{Duk}, obtained by multiplying the
lowest-order Drell-Yan calculations by the correction factor
$K^{(i)}$.  No further multiplicative factors are needed to obtain
the very good fit shown in fig.~\ref{fig1}, which thus
demonstrates the reliability of the correction factor
$K^{(i)}$. It is interesting to note that the function $K^{(i)}$
from Eq.~(\ref{eqn:DYK}) which gives a good fit in Fig.~\ref{fig1}
is not a constant of the invariant mass $M$. It is equal to $2.5$
for $M=3.9$ GeV and $1.8$ for $M=17$ GeV. There is thus a $30$ \%
variation of $K^{(i)}$ as $M$ varies from $4$ to $17$ GeV.

For lowest-order calculations, the rate for the production of
dileptons with an invariant mass $M$ per unit four-volume in a
thermalized quark-gluon plasma (with three flavors) depends on the
temperature $T$ of the system and can be written in the form
\cite{Kaj2,Won1}
\begin{eqnarray}
  {dN_{l^+l^-} \over dM^2 d^4x} \approx N_c N_s^2
  \sum_{q=u,d,s} \biggl ({e_{q} \over e} \biggr )^2 {\sigma_q
    (M) \over 2(2\pi)^4} M^2 \sqrt{ 1-{4m_q^2 \over M^2} }\; T M
  K_1\biggl ({M\over T} \biggr ) \,,
\label{eq:dndq}
\end{eqnarray}
where $\sigma_q(M)$ is the lowest-order $q\bar q \rightarrow
l^+ l^-$ cross section at the center-of-mass energy $M$
given by \cite{Won1}
\begin{eqnarray}
  \sigma_q(M) = {4 \pi \over 3} { \alpha^2 \over M^2} \biggl (1
  - {4 m_q^2 \over M^2} \biggr )^{\!\!-{1\over2}} \!\!\sqrt{
    1-{4m_l^2 \over M^2}} \biggl ( 1+ 2{ m_q^2+m_l^2 \over M^2}
  +4{m_q^2 m_l^2 \over M^4} \biggr ) \,,
\end{eqnarray}
$m_l$ and $m_q$ are the rest masses of the lepton $l$ and the quark
$q$ respectively, and $K_1$ is the modified Bessel function of first
order.

The integration of Eq.~(\ref{eq:dndq}) over the transverse
dimensions and the proper time interval $\tau_0$ (the
formation time) to $\tau_c$ (the hadronization time) gives the
invariant mass-squared distribution of dileptons produced from
the system while it was in the quark-gluon plasma phase as
\cite{Kaj2,Won1}
\begin{eqnarray}
  {dN_{l^+l^-} \over dM^2 dy} \approx {\cal A} N_c N_s^2
  \sum_{q=u,d,s} \biggl( {e_q \over e}\biggr)^2 {\sigma_q(M)
    \over 2(2\pi)^4} \biggl(1-{4m_q^2 \over M^2}\biggr)^{1 \over
    2}\; {3 \tau_0^2 T_0^6 \over M^2 } \biggl[ H\biggl({M\over
    T_0}\biggr) - H\biggl({M\over T_c}\biggr) \biggr] \,,
\label{eq:dndq2}
\end{eqnarray}
where ${\cal A}$ is the transverse area of the collision region,
$T_0$ and $T_c$ are the initial and final temperatures of the
ideal QGP phase, and

\begin{eqnarray*}
  H(z)=z^2 (8+z^2)K_0(z)+4z(4+z^2)K_1(z)\,.
\end{eqnarray*}

To delineate the QGP dileptons from other sources of dileptons,
particularly the Drell-Yan background, we need to have good estimates
of the dilepton production from the different sources at matching
levels of accuracy.  It is hence necessary to modify
Eq.~(\ref{eq:dndq2}) by incorporating the higher-order QCD
corrections.  The interpolation from low to high relative velocities
extends the usefulness of the correction factors $K^{(i)}$ given by
Eq.~({\ref{eq:150}) to account for higher-order QCD effects. The
  modified rate of dilepton production then becomes
\begin{eqnarray}
  {dN_{l^+l^-} \over dM^2 dy} \approx {\cal A} N_c N_s^2
  \sum_{q=u,d,s} K^{(i)}(q ) \biggl ({e_{q} \over e} \biggr
  )^2 { \sigma_q(M) \over 2(2\pi)^4} \biggl(1-{4m_q^2 \over
    M^2}\biggr)^{1 \over 2}\; {3 \tau_0^2 T_0^6 \over M^2 }
  \biggl[ H\biggl({M\over T_0}\biggr) - H\biggl({M\over
    T_c}\biggr) \biggr] \,.
\label{eq:dndq2k}
\end{eqnarray}

To calculate Eq.~(\ref{eq:dndq2}) or Eq.~(\ref{eq:dndq2k}), we
need to know the temperatures $T_0$ and $T_c$, and the
``formation'' or ``materialization'' time $\tau_0$. We use
$T_c=180$ MeV.  The Monte Carlo simulation program ``MARCO''
\cite{Won2}, which is based on the Glauber multiple-collision
model and reproduces the peak value of $dN/dy$ in nucleus-nucleus
collisions, is used to calculate the rapidity density of the
produced particles (mostly pions).  The peak energy density
${\bf\varepsilon}$ is related to the peak rapidity density
$dN/dy|_{peak}$ in the Bjorken scaling hydrodynamic
model\cite{Bjo,Won1} as
\begin{equation}
  \label{eq:energy} {\bf\varepsilon} = {m_{_T} \over \tau_0
  \cal{A}} \; {dN \over dy}\biggr|_{peak}\,,
\end{equation}
where $m_{_T}= \sqrt{m_\pi^2+p^2_{_T}}$, $m_\pi$ is the pion mass,
and $p_{_T}$ is the pion transverse momentum, which varies with
the center-of-mass energy of the colliding nuclei as\cite{Won1}
\begin{eqnarray*}
  p_{_T}= 0.27+0.037 \ln(\sqrt{s})\,.
\end{eqnarray*}
The temperature $T_0$ is then approximated as\cite{Bjo,Won1}
\begin{equation}
  \label{eq:temp}
  T_0 =\; \Bigl[{30\over 37 }\;{{\bf\varepsilon} \over \pi^2 }
  \Bigr]^{1/4}\,.
\end{equation}
The value of the formation time $\tau_0$ is found to play an
important role in determining the value of the initial temperature
$T_0$ of the QGP. It has been given different values in the
literature.  In the following, we use the values obtained from
MARCO for the peak $dN/dy$, typically $2144$ and $231$, in the
central collisions of Au + Au and S + S, respectively, at RHIC
energies.  The common argument assumes that $\tau_0$ should be of
the order of a typical strong interaction time scale of $1.0$
fm/c, as proposed by Bjorken\cite{Bjo}.  Using this value,
Eq.~(\ref{eq:energy}) and Eq.~(\ref{eq:temp}) then give $T_0=262$
MeV for Au+Au and $T_0=202$ MeV for S+S.  Shuryak {\em et al.}
\cite{Shu93,Shu94} assume a different scenario for calculating the
thermalization time $\tau_0$. In the ``hot glue'' scenario, the
thermalization time of the quark-gluon plasma is expected to be
within the range $0<\tau_0<0.3$ due to gluon rescattering and
production mechanisms. This picture gives a temperature much
higher than that of the Bjorken scenario.  Ruuskanen\cite{Ruu}
assumes that the quark-gluon plasma may thermalize at an early
time $\tau_0=0.5$ fm/c which gives $T_0=312$ MeV for Au+Au and
$T_0=241$ MeV for S+S.  Alternatively, Kapusta\cite{Kap} uses a
different approach to estimate $\tau_0$. Assuming that the partons
will have a thermal distribution right after they are put on the
mass shell, and that the initial rapidity density is the same as
the observed one, Kapusta concludes that $\tau_0\/T_0=constant$.
Using the uncertainty principle and putting $\Delta E=3\/T_0$, he
then estimates the $constant$ to be $1/3$, i.e
\begin{equation}
  \label{eq:tau}
  \tau_0\/T_0 = 1/3\,,
\end{equation}
which gives $T_0=590$ MeV and $\tau_0=0.11$ fm/c for the collision
Au+Au and $T_0=421$ MeV and $\tau_0=0.16$ fm/c for the collision
S+S.

Due to the importance of the charm dilepton background, it is necessary
to estimate the charm contribution for a comparitive study. For charm
production, the gluon fusion channels as well as the $q\bar q$
annihilation via virtual gluon modes must be included\cite{Com}.

The basic cross section for $q\bar q \rightarrow g^* \rightarrow c\bar c$,
averaged over initial and summed over final colors and spins can be
written as
\begin{eqnarray}
\sigma_{q\bar q} (M_{c\bar c}) =   { 8 \pi\alpha^2_s \over 27 M_{c\bar
c}^2}
\biggl (1+{\eta
\over 2}\biggr) \sqrt{1-\eta}
\end{eqnarray}
where $\eta =4 m^2_c/M_{c\bar c}^2.$, with $m_c$ being the mass of
the charm quark and $M_{c\bar c}$ being the invariant mass of the
produced $c\bar c$ pair.

The corresponding expression for the gluon fusion mode, averaged
over initial gluon types and polarizations and summed over final
colors and spins is \cite{Won1}
\begin{eqnarray}
\sigma_{gg} (M_{c\bar c}) = { \pi\alpha^2_s \over 3 M_{c\bar
c}^2}\biggl\{ (1 +\eta +{1\over 16} \eta^2) \ln
\biggl({1+\sqrt{1-\eta}\over 1-\sqrt{1-\eta}} \biggr) - \biggl(
{7\over 4} + {31\over16} \eta \biggr) \sqrt{1-\eta}
\biggr)\biggr\}
\end{eqnarray}
The above cross sections must be convoluted over the quark and
gluon distributions respectively as for the DY case to obtain the
overall charm yield for comparison with the QGP signals. The charm
production can then be written for the nucleon-nucleon case as
\begin{eqnarray}
  {d^2\sigma^{NN}_{c\bar c} \over dM_{c\bar c}dy_{c\bar c}}&=&K
 \biggl\{ \biggl( \sum_{q=u,d,s} \sigma_{q\bar q}(M_{c\bar c}) [
 q_q^a (x e^y) \bar q_q^b (x e^{-y}) + \bar q_q^a (x e^y)q_q^b (x
 e^{-y})] \biggr ) \nonumber\\ & & + \biggl([ g^a(x e^y)g^b(x
 e^{-y})]+ [a \leftrightarrow b ] \biggr) \sigma_{gg}(M_{c\bar c})
 \biggr \}
\end{eqnarray}
where $x=M_{c\bar c}/ \sqrt{s}$, $M_{c\bar c}$, $y_{c\bar c}$ are
the invariant mass and the rapidity of the $c\bar c$ pair.  The
factor $K$ in this case is an empirical factor and is usually
taken to be $3$\cite{Ber}.

    The $c\bar c$ pairs produced by the $q\bar q$ or $gg$
reactions hadronize to $D$ mesons which subsequently decay through
semi-leptonic channels, such as $D \rightarrow l + X $ to
leptons. Leptons from a decaying $D\bar D$ pair comprise a
dilepton and the charm production rates can be mapped into the
resulting decay dilepton distributions with invariant mass
$M_{l+l-}$ and rapidity $y_{l+l-}$ by a Monte Carlo technique to
generate the dileptons from charm decays.

The nuclear-nuclear charm dilepton cross section can be obtained
from the nucleon-nucleon one according to

\begin{eqnarray}
  \label{eq:AADD} {d^2N^{AA}_{l^+ l^-} \over dM_{l^+ l^-}dy_{l^+
    l^-}} = {3 \over 4 \pi (r'_0)^2}\; A^{4/3}\; {d^2
    \sigma^{NN}_{l^+ l^-} \over dM_{l^+ l^-}dy_{l^+ l^-}} \,,
\end{eqnarray}

The dilepton production rates per unit time from a thermalized
QGP, which may be formed after the collisions of Au+Au and S+S at
an energy $\sqrt{s}=200$ GeV, for three different values of the
variable $\tau_0$ are shown in Fig.~\ref{fig2} and
Fig.~\ref{fig3}, respectively. The solid curves in
Fig.~\ref{fig2}(a) and Fig.~\ref{fig3}(a) show those rates given
by Eq.~(\ref{eq:dndq2}) using the results of the lowest-order
cross section of dileptons production. The solid curves in
Fig.~\ref{fig2}(b) and Fig.~\ref{fig3}(b) show the same rates
taking into account the correction factor $K^{(i)}$
[Eq.~(\ref{eq:dndq2k})].  The dotted-dash curves in both
Fig.~\ref{fig2} and Fig.~\ref{fig3} show the Drell-Yan rates.  The
dotted curve is the dilepton rates from the open-charm mesons
decays.

The virtual photon annihilation mode selects the annihilating
$q\bar q$ pair to be in the color-singlet states and thus the
interaction between $q$ and $\bar q$, before they annihilate, is
attractive.  The effect of $q \bar q$ initial-state interaction
leads to an enhancement factor $K^{(i)}$, as shown in
Fig.~\ref{fig2} and Fig.~\ref{fig3}.  Because the coupling
constant increases and the relative velocity decreases as the
invariant mass decreases, the correction factor rises considerably
with the decrease of invariant mass.  The effect of the
higher-order QCD corrections is to enhance the tree-level dilepton
cross section by a factor of about 5 at $\sqrt{s}=1$ GeV, and by a
factor of about 3 at $\sqrt{s}=2-3$ GeV.

It is seen that for the collision Au+Au, the temperatures
$T_0=262$ MeV and $T_0=312$ MeV lead to the result that the
dilepton rates from the Drell-Yan process (the dashed curves in
Fig.~\ref{fig2}) exceed those from the QGP (the solid curves in
Fig.~\ref{fig2}(a)) above $1$ GeV, and the charm dilepton
contribution is even higher. The temperature $T_0=590$ Mev allows
the uncorrected thermal rate to exceed the DY and charm dileptons
up to almost $6$ Gev and indicates it may be possibile to detect
the signal. However, the inclusion of the correction factor
$K^{(i)}$ changes the situation dramatically. It increases the
thermal rates (solid curves in Fig.~\ref{fig2}(b)) so that they exceed the
Drell-Yan background up to an invariant mass of $M=2$ GeV even for
the lower temperature models and is comparable to the charm
dileptons, which may prove to be detectable if the contribution of
the charm dileptons can be found by independent charm measurement.
For the temperature $T_0=590$ MeV, it further improves the
prospects to detect the QGP dileptons since their rates are now
well above the background up to $M=7.5$ GeV.

Similarly, for the collision S+S at an energy $\sqrt{s}=200$ GeV,
as shown in Fig.~\ref{fig3} at lower temperatures $T_0=202$ MeV
and $T_0=241$ MeV, the dilepton rates from the Drell-Yan and charm
background exceed those from the QGP. The inclusion of the
correction factor $K^{(i)}$ enances the QGP dilepton yield so that it exceeds
the Drell-Yan dilepton yield up to an invariant mass of $M=1.5$ GeV, and
is comparable to the charm dileptons. For the temperature
$T_0=421$ MeV, the QGP dilepton rate, without the correction
factor $K^{(i)}$, exceeds the Drell-Yan dileptons up to an
invariant mass of $M=4.5$ GeV, and the charm dileptons up to $2$
GeV.  With the correction factor $K^{(i)}$, the thermal dilepton
yield exceeds the DY yield up to $5.5$ GeV and the charm dileptons
up to $3.5$ GeV, so that detection prospects increase considerably.

While different temperatures have been proposed for the initial
state of the QGP, the observation of the initial temperature is an
experimental question. In this regard, higher-order QCD
corrections, as accounted for by using the $K^{(i)}$ factor,
enhance the possibility of detecting dileptons from the
quark-gluon plasma for both Au+Au and S+S collisions, even if the
initial temperature turns out to be lower than about $350$ MeV. On
the other hand, without the QCD corrections, the thermal dilepton
yield from the QGP is lower than that from the Drell-Yan and charm
decay processes and will be difficult to detect. It may be noted
that recent calculations, approaching the first order corrections
in a different manner, using thermal masses and finite temperature
QCD effects also indicate an enhanced dilepton yield from the
thermal plasma\cite{Alt}.

In the plasma, the color charge of the constituents is subject to
Debye screening which is characterized by the Debye screening
length $\lambda_D$, which depends on the temperature $T$. The
Debye screening length at a temperature of $400$ MeV is about
$\lambda_D \sim 0.2$ fm.  On the other hand, the electromagnetic
annihilation into dileptons occurs within an extremely collapsed
space zone characterized by the linear $q$-$\bar q$ distance $\sim
\alpha/\sqrt{s}$, which is about $0.0029$ fm for the annihilation
of a light quark-antiquark pair at 0.5 GeV, and is much smaller
than the Debye screening length $\lambda_D$. The interaction
between the quark and the antiquark is not expected to be much
affected by Debye screening in the region where annihilation
occurs.  Therefore, we expect that our correction factor for
dilepton production by $q \bar q$ annihilation in the plasma will
not be modified much by the addition of Debye screening
corrections.

 In the charm estimates, the open-charm production was based on a
model of gluon fusion and $q\bar q$ annihilation, with an
empirical $K$ factor independent of the color state of the
reacting partons.
In this region of $c\bar c$ production near the threshold, it is
expected that the initial- and final-state interactions of the
participating gluons, quarks, or antiquarks are important, and the
effective interaction depends on the color multiplet of the
participating partons\cite{Lal}. A careful examination of these
known effects to study open-charm production will be of interest
to augment the investigation here and to reexamine the results of
\cite{Aki}.

    Although we have specifically focussed on the experimental
situation expected at RHIC, the use of the correction factor is
not restricted to this. In fact, the use of the $K^{(i)}$ can be
extended to any experiment designed to investigate the QGP
formation or existing data and it can be applied to account for
the higher order QCD corrections to the lowest order thermal
dilepton production in a compact way. In all situations where an
intial distribution of the quarks and anti-quarks in the thermal
plasma can be assumed, the factor can be used to provide estimates
of the thermal dilepton production, including the higher order
corrections and thus can serve as an useful tool in determining
QGP signatures.

\acknowledgements

One of the authors (M.M.) would like to thank Prof. A. Goneid for
helpful discussions and Ain Shams University for financial
support.  L. C. would like to thank the University Grants
Commission of India for partial support.  M. M., L. C. and
Z.W. would like to thank Dr. M.  Strayer, Dr. J. Ball and
Dr. F. Plasil for their kind hospitality at ORNL.  This research
was supported in part by the Division of Nuclear Physics,
U.S. Department of Energy under Contract No. DE-AC05-84OR21400
managed by Martin Marietta Energy Systems, Inc.

\begin{figure}[htbp]
 \caption{The differential cross section $s d^2\protect\sigma/d
   \protect\sqrt\tau dy$ as a function of $\protect\sqrt\tau$ for
   different rapidity $y$ intervals. The data are taken from the
   FANL-605 experiment \protect\cite{Sti93}. The solid curve are
   the fitting of the lowest order QCD calculations multiplied by
   the correction factor $K^{(i,f)}$, see text.}  \label{fig1}
\end{figure}

\begin{figure}[htbp]
 \caption{Dilepton production rates from the collision Au+Au at
   RHIC energies ($\protect\sqrt{s}=200$ GeV) and $dN/dy=2144$:
   (a) without the correction factor $K^{(i)}$ and (b) with the
   correction factor $K^{(i)}$.}  \label{fig2}
\end{figure}

\begin{figure}[htbp]
 \caption{Dilepton production rates in the collision S+S at RHIC
   energies ($\protect\sqrt{s}=200$ GeV) and $dN/dy=231$: (a)
   without the correction factor $K^{(i)}$ and (b) with the
   correction factor $K^{(i)}$.}  \label{fig3}
\end{figure}

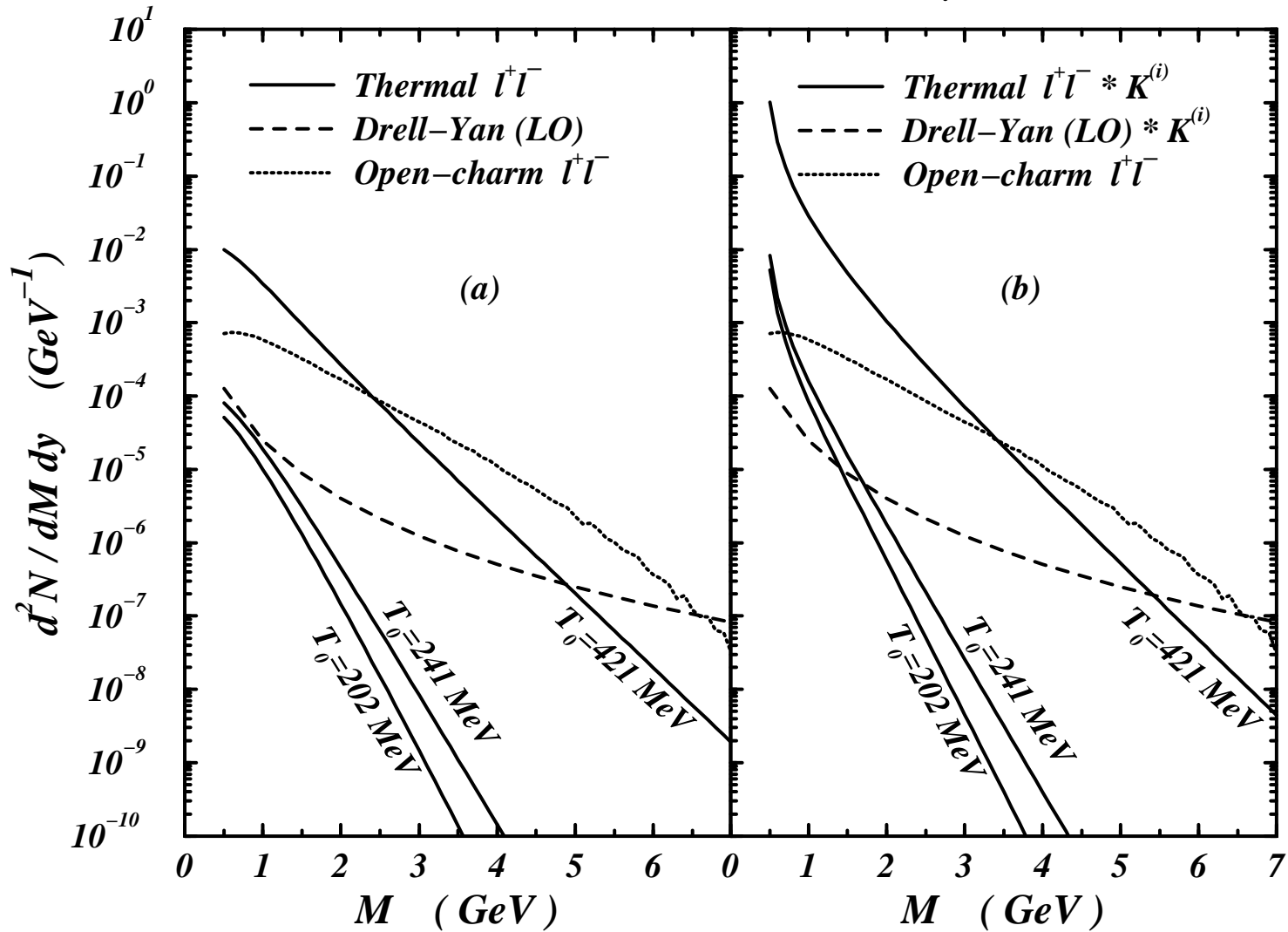
\end{document}